\documentclass[%
 preprint,
 amsmath,amssymb,
 aps, physrev,
]{revtex4-2}

\usepackage{graphicx}
\usepackage{dcolumn}
\usepackage{bm}
\usepackage[hidelinks]{hyperref}
\usepackage{academicons}
\usepackage{xcolor}
\usepackage{orcidlink}
\usepackage{lipsum}
\usepackage{natbib}

\usepackage[nameinlink,noabbrev]{cleveref}
\crefname{equation}{Eq.}{Eqs.}
\Crefname{equation}{Equation}{Equations}
\crefname{figure}{Fig.}{Figs.}
\Crefname{figure}{Figure}{Figures}
\newcommand{\A}{\operatorname{A}}
\newcommand{\Ma}{\mathrm{M}}

\newcommand{\T}{\mathsf{T}}
\newcommand{\mix}{\text{m}}
\newcommand{\Ev}{\operatorname{Ev}}
\newcommand{\tv}{\text{v}}
\newcommand{\tc}{\text{c}}

\newcommand{\Real}{\operatorname{Re}}
\usepackage{siunitx}
\usepackage{comment}
\usepackage{float}
\usepackage{tabularx}
\usepackage{array}

\begin{document}

\preprint{APS/123-QED}

\title{\textbf{Early-time dynamics of the Richtmyer--Meshkov instability at Kelvin--Voigt viscoelastic interfaces}}

\author{Xuning Zhao\orcidlink{0000-0001-8019-0999}}
\author{Mauro Rodriguez Jr.\orcidlink{0000-0003-0545-0265}}\email{Contact author: mauro\_rodriguez@brown.edu}
\affiliation{%
School of Engineering, Brown University, Providence, RI
}%

\date{\today}

\begin{abstract}
The Richtmyer–Meshkov instability (RMI) is initiated when a shock wave interacts with a perturbed interface between two materials, imparting an initial velocity to the interfacial perturbation.
When the materials are viscoelastic, the subsequent evolution is influenced by the coupled effects of inertia, viscous dissipation, and elastic resistance.
The interplay of these effects remains poorly understood and is important for predicting the dynamic response of shock-loaded soft materials, e.g., in blast-induced injury.
Using the incompressible Cauchy momentum equation with a Kelvin--Voigt constitutive model, we conduct a linear stability analysis for the early-time post-shock evolution of a single-mode interface perturbation of two viscoelastic materials.
We derive an implicit dispersion relation for the temporal growth rates of the perturbation.
This relation depends on density, viscosity, and elasticity contrasts between the two materials and an viscoelastic parameter which is a bounded ratio of the viscous dissipation and elastic resistance response rates.
Four material cases, spanning symmetric to highly contrasted interfaces, are examined to show how the viscoelastic number affects the growth rate structure as complex-dominated (oscillatory) or real-dominated (monotonic relaxation).
The critical viscoelastic number separates these two regimes and is primarily governed by the density and viscosity contrasts.
These results provide a theoretical criterion for predicting the early-time response of shock-driven interfaces involving fluids and viscoelastic solids.

\end{abstract}

\maketitle

\section{Introduction\label{sec:introduction}}

The Richtmyer--Meshkov instability (RMI) arises when a shock wave interacts with a perturbed interface separating two materials of different densities~\cite{Richtmyer1960,Meshkov1969,brouillette2002richtmyer}.
It plays an important role in engineering systems including supernova explosions~\cite{Abarzhi2019}, supersonic combustion~\cite{Chou2024}, and the dynamic response of shock-loaded materials~\cite{Vogler2021}.
The RMI process is commonly decomposed into a shock-interface interaction stage followed by a post-shock evolution stage~\cite{brouillette2002richtmyer}.
During shock passage, the misalignment of pressure and density gradients deposits baroclinic vorticity at the interface, initiating the subsequent post-shock interface evolution.
The post-shock evolution is then governed by the interplay of inertia and the material constitutive response.
The resulting interface deformation may develop into complex interfacial structures and promote mixing between the two materials~\cite{Prestridge2013,Zhou2021,Zhou2017a}.

Classical descriptions of RMI generally treat the materials as inviscid fluids. 
For example, Richtmyer's impulsive model represents the shock as an instantaneous acceleration and relates the initial post-shock perturbation growth rate to the perturbation amplitude and wavenumber, density contrast, and shock-induced velocity impulse~\cite{Richtmyer1960}. 
Subsequent studies extended the classical description to account for compressible wave dynamics~\cite{Fraley1986,Wouchuk1996}, finite-amplitude and nonlinear perturbation growth~\cite{Velikovich1996}, and interactions among multiple perturbation modes~\cite{Gauthier2002}.
However, when the material response involves viscous dissipation or elastic resistance, the post-shock interface evolution can be significantly modified.
For fluid interfaces, Newtonian viscosity dissipates interfacial motion and attenuates interface perturbations~\cite{mikaelian1993,carles2002,mikaelian2019}.
On the other hand, elasticity in solid materials resists deformation, providing a restoring force that can bound the perturbation amplitude or produce oscillatory interface motion~\cite{sun2018effects,Plohr2005,Piriz2006,LopezOrtega2010}.

Soft materials, including hydrogels, polymeric gels, and biological tissues, exhibit both  viscous and elastic responses~\cite{Chaudhuri2016,Lou2021,Chaudhuri2020,Kobayashi2017}.
When an interface involving such materials is subjected to shock loading, its evolution reflects the coupled influence of viscous dissipation and elastic resistance.
Such interfacial dynamics constitute an emerging problem with potential relevance to heterogeneous soft materials subjected to impulsive loading.
For example, in blast-induced lung injury, rapid loading of the numerous air--tissue interfaces may lead to alveolar instability, which can damage the alveolar--capillary structure~\cite{Steinberg2004}.
In blast-induced brain injury, mechanical and acoustic impedance mismatch among the skull, cerebrospinal fluid, and brain tissues concentrate pressure and shear stresses to potentially produce localized tissue deformation and damage~\cite{Nakagawa2011,Park2011}.
These settings motivate a better understanding of shock-driven interfacial deformation in viscoelastic materials.

Recent experiments have reported the development of shock-driven interfacial instabilities in soft hydrogels and viscoelastic materials~\cite{Deng2025}.
These experiments revealed post-shock interfacial oscillations in soft hydrogels, a behavior not observed in fluid interfaces.
Using a nonlinear potential-flow model for an interface separating two Rivlin--Ericksen viscoelastic fluids, ~\cite{Rollin2011} showed that viscoelasticity prolongs the late-time interface evolution by slowing the decay of the interface velocity.
More recently, a linear model based on the Maxwell constitutive relation was proposed for the early-time evolution of RMI at gas--viscoelastic interfaces and experimentally reproduced the oscillatory perturbation response~\cite{Deng2025}.
However, it remains unclear under what material parameter configurations does the interface evolves monotonically (viscous-dominated) or oscillates (elasticity-dominated).

The aim of this work is characterize the early-time post-shock evolution of the material interface separating two viscoelatic materials.
We assume the fluid flow is in the low-Mach-number regime (i.e., incompressible) and use linear theory to analyze the fluid dynamics.
The two materials are modeled using the Kelvin--Voigt constitutive relation.
This relation is widely used for small-strain mechanical response of soft biological tissues, hydrogels, and other soft polymeric materials~\cite{Parker2019,Zhang2021,Li2021,Klatt2007,Weickenmeier2018,DeGeeter2009,Jamburidze2017,Shabaniverki2017,Sharma2020,sanchez2026hierarchical}.
We derive a dispersion relation to identify the admissible temporal growth rate structures and characterize their dependence on the viscoelastic material properties.
The paper is organized as follows. 
\Cref{sec:problem_setup} presents the problem formulation, governing equations, and constitutive model.
\Cref{sec:post-shock} shows the derivation and analysis of the dispersion relation governing the early-time post-shock evolution.
\Cref{sec:results} presents the resulting growth rate structures for representative material configurations and examines the effects of density, viscosity, and elasticity on the interface dynamics.
In \Cref{sec:summary}, we summarizes our findings.

\section{Problem setup and governing equations\label{sec:problem_setup}}
\begin{figure*}
    \centering
    \includegraphics[width=\linewidth]{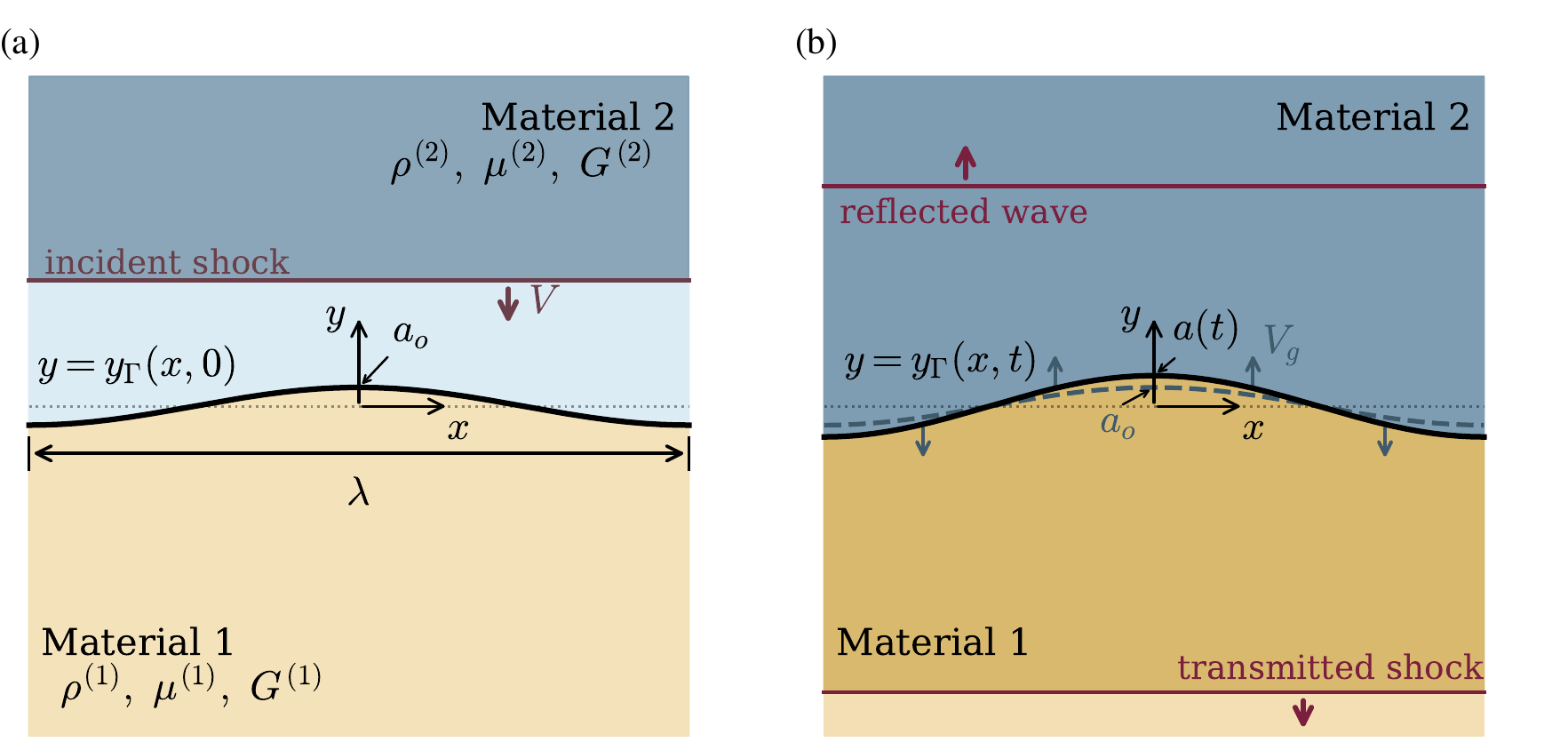}
    \caption{
    Configuration schematic for the Richtmyer--Meshkov instability between viscoelastic materials.
    (a) An incident shock approaches a perturbed viscoelastic interface; (b) post-shock evolution of the interface perturbation in the linear growth regime.
    } 
    \label{fig:problem_illustration}
\end{figure*}

\Cref{fig:problem_illustration} shows the problem setup in the initial and post-shock configurations.
The problem is formulated in a rectangular Cartesian coordinate system.
Two immiscible viscoelastic fluids are separated by an interface, with the lower and upper fluids occupying semi-infinite domains as $y\to\pm\infty$, respectively.
The interface is initialized with a small-amplitude single-mode perturbation of amplitude $a(t=0) = a_o$ and wavenumber $k$.
In the $x$-direction, the domain spans one wavelength with $\lambda=2\pi/k$ and has periodic boundary conditions.
The interface is described by
\begin{equation}
    \mathbf{x}_{\Gamma} = 
    \left[\begin{array}{c}
         x_{\Gamma}\\
         y_{\Gamma}
    \end{array}\right] = 
    \left[\begin{array}{c}
         x_{\Gamma}\\
         a(t)\cos(kx)
    \end{array}\right],
\end{equation}
where $a(t)$ is time-dependent perturbation amplitude. 
The materials have density $\rho^{(j)}$, shear viscosity $\mu^{(j)}$, shear modulus $G^{(j)}$, and base pressures $p^{(j)}$, where superscript $j$ are $1$ and $2$ for the lower fluid ($y<y_{\Gamma}(x,0)$) and upper fluid ($y>y_{\Gamma}(x,0)$), respectively.

A planar shock wave travels towards the interface with speed $V$.
As the shock traverses the perturbed interface, impulsive loading deposits baroclinic vorticity at the interface, thereby initiating the subsequent RMI.
Both the inital and post-shock perturbation amplitude are assumed to satisfy $k a \ll1$, such that the interface remains weakly perturbed.
The characteristic timescales associated with shock traversal across the interface and the subsequent interface evolution are $\tau \sim \lambda/V$ and $\tau_{\text{g}} \sim \lambda/V_{\text{g}}$, respectively, where $V_{\text{g}}$ denotes the characteristic interface growth velocity.
Since $V_{\text{g}} \ll V$, the corresponding timescales satisfy $\tau \ll \tau_{\text{g}}$.
Consequently, the shock completely traverses the interface before substantial post-shock interface evolution occurs.
The interface evolution may then be viewed as a rapid compressible shock-interaction stage followed by a slower post-shock instability growth stage.
The post-shock evolution is assumed to be isothermal, such that the viscosity and shear modulus remain constant throughout the subsequent interface evolution.
Moreover, for soft viscoelastic materials, the acoustic wave speed scales as $c\sim\sqrt{K/\rho}$, where $K$ is the bulk modulus. 
As a representative example, hydrogels typically possess bulk moduli on the order of $K\sim\SI{1}{\giga\pascal}$, corresponding to acoustic wave speeds of order $\SI{E3}{\meter\per\second}$~\cite{Pickard2023}.
The post-shock fluids are assumed to remain in the low-Mach-number regime ($\Ma=V_{\mathrm{g}}/c<0.3$), where characteristic velocity remains small compared to the acoustic velocity scale, and therefore modeled as incompressible.

The governing equations are continuity and linear momentum balance:
\begin{subequations}
    \label{eq:governing}
    \begin{equation}
        \boldsymbol{\nabla} \cdot \mathbf{v}^{(j)} = 0,
        \label{eq:continuity}
    \end{equation}
    \begin{equation}
        \rho^{(j)}\frac{D\mathbf{v}^{(j)}}{Dt} =  \boldsymbol{\nabla}\cdot\boldsymbol{\sigma}^{(j)},
        \label{eq:momentum_viscous_hyper}
    \end{equation}
\end{subequations}    
where superscript $j=1,2$ indicates the material domain (bottom and top, respectively), $\mathbf{v}^{(j)}$ is the velocity vector, and $D/Dt$ the material derivative.
The Cauchy stress tensor is
\begin{equation}
    \boldsymbol{\sigma}^{(j)} = -p^{(j)}\mathbf{I} + \boldsymbol{\tau}^{(j)},
\end{equation}
where $p^{(j)}$ is the pressure, $\mathbf{I}$ the identity tensor, and $\boldsymbol{\tau}^{(j)}$ the deviatoric stress contribution.
For the latter stress contribution, the Kelvin--Voigt constitutive model is used and consists of a Newtonian viscous element and linear Hookean spring connected in parallel:
\begin{equation}
    \boldsymbol{\tau}^{(j)} = \boldsymbol{\tau}^{(j)}_{\tv} + \boldsymbol{\tau}^{(j)}_{\text{e}},
    \label{eq:constitutive_law}
\end{equation}
where $\boldsymbol{\tau}^{(j)}_{\tv}$ and $\boldsymbol{\tau}^{(j)}_{\text{e}}$ are the respective viscous and elastic terms:
\begin{subequations}
\begin{equation}
    \boldsymbol{\tau}^{(j)}_{\tv} = \mu^{(j)} \left[\boldsymbol{\nabla}\mathbf{v}^{(j)}+(\boldsymbol{\nabla}\mathbf{v}^{(j)})^\T\right], 
    \label{eq:cauchy_viscous}
\end{equation}
\begin{equation}
    \boldsymbol{\tau}^{(j)}_{\text{e}} = G^{(j)} \left[\boldsymbol{\nabla}\mathbf{u}^{(j)}+(\boldsymbol{\nabla}\mathbf{u}^{(j)})^\T\right],
\end{equation}
\end{subequations}
with $\mu^{(j)}$ as the shear viscosity,  $G^{(j)}$ the shear modulus, and $\mathbf{u}^{(j)}$ the deformation vector. 

\Cref{eq:governing} are coupled at $\mathbf{x}_{\Gamma}$ by no-slip, no-penetration, and continuity of stress traction conditions across the interface, respectively:
\begin{subequations}
\label{eq:interface_conditions}
    \begin{equation}
        (\mathbf{v}^{(1)} - \mathbf{v}^{(2)})\cdot \hat{\mathbf{n}}=0,
    \end{equation}
    \begin{equation}
        (\mathbf{v}^{(1)} - \mathbf{v}^{(2)})\cdot \hat{\mathbf{t}}=0,
    \end{equation}
    \begin{equation}
       \mathbf{\hat{n}} \cdot (\boldsymbol{\sigma}^{(1)} - \boldsymbol{\sigma}^{(2)})\mathbf{\hat{n}} = 0,
    \end{equation}
    \begin{equation}
        \mathbf{\hat{t}}\cdot\left(\boldsymbol{\sigma}^{(1)} - \boldsymbol{\sigma}^{(2)}\right)\mathbf{\hat{n}}  = 0,
    \end{equation}
\end{subequations}
where $\hat{\mathbf{n}}$ and $\hat{\mathbf{t}}$ are the normal and tangential unit vectors at $\mathbf{x}_{\Gamma}$.

\section{Post-shock Perturbation Analysis}
\label{sec:post-shock}

\subsection{Method of normal modes}
\label{sec:normal_modes}

The flow variables in each material domain are decomposed as
\begin{equation}
    \phi^{(j)} = \phi_{\text{s}}^{(j)} + \epsilon \phi_1^{(j)},
    \label{eq:states_linearized}
\end{equation}
where the subscript s is the post-shock base state and $1$ the first-order perturbation state with amplitude $\epsilon \ll 1$.
Linearizing~\cref{eq:governing} and retaining the first-order perturbation equations:
\begin{subequations}
\label{eq:linearized_eqs}
    \begin{equation}
        \boldsymbol{\nabla}\cdot\mathbf v_1^{(j)} = 0,
        \label{eq:linearized_continuity}
    \end{equation}
    \begin{equation}
        \rho_{\text{s}}^{(j)} \frac{\partial \mathbf{v}_1^{(j)}}{\partial t} =  -\boldsymbol{\nabla} p_1^{(j)} + \mu^{(j)}\boldsymbol{\nabla}^2\mathbf v_1^{(j)}  + G^{(j)}\boldsymbol{\nabla}^2\mathbf u_1^{(j)}.
        \label{eq:linearized_momentum}
    \end{equation}
\end{subequations}

The method of normal modes is used to define the perturbation states of the velocity, displacement, and pressure fields:
\begin{equation}
    \phi_1^{(j)}(x,y,t) = \Phi^{(j)}(y) e^{ikx + \gamma t},
    \label{eq:normalmode}
\end{equation}
where capital field quantities $\Phi(y)$ are the amplitudes of $\phi$, $i = \sqrt{-1}$ is the imaginary root, and $\gamma$ the temporal growth rate.
Since $\mathbf{v}^{(j)} = \partial \mathbf{u}^{(j)}/\partial t$ in the linear limit, the modal amplitudes satisfy $\mathbf{V}^{(j)}=\gamma\mathbf{U}^{(j)}$.
Substituting the ansatz into \cref{eq:linearized_eqs} yields
\begin{subequations}
    \begin{equation}
        ikV^{(j)}_x + D V_y^{(j)} = 0,
        \label{eq:normal_continuity}
    \end{equation}
    \begin{equation}
        \rho_{\text{s}}^{(j)} \gamma V_{x}^{(j)} = -ik P^{(j)} + \hat{\mu}^{(j)} (D^2 - k^2) V_x^{(j)},
        \label{eq:xnormal}
    \end{equation}
    \begin{equation}
        \rho_{\text{s}}^{(j)} \gamma V_y^{(j)} = - D P^{(j)} + \hat{\mu}^{(j)} (D^2 - k^2) V_y^{(j)},
        \label{eq:ynormal}
    \end{equation}
    \label{eq:normal_momentum}%
\end{subequations}
where subscripts $x$ and $y$ denote the $x$- and $y$-direction, respectively, $V_x^{(j)}$ and $V_y^{(j)}$ are the modal velocity vector amplitudes, and $D \equiv d/dy$.
The growth-rate-dependent effective viscosity includes an elastic contribution:
\begin{equation*}
    \hat{\mu}^{(j)} = \mu^{(j)} + \frac{G^{(j)}}{\gamma}.
\end{equation*}

Combining~\cref{eq:normal_continuity,eq:xnormal,eq:ynormal} and eliminating $V_x^{(j)}$ and $P^{(j)}$ yields the velocity eigenfunction in the $y$-direction:
\begin{equation}
    (D^2 - k^2)\left(D^2 - k^2 - \rho_{\text{s}}^{(j)} \frac{\gamma}{\hat\mu^{(j)}} \right)V_y^{(j)} = 0.
    \label{eq:V_eigenfunction}
\end{equation}
This equation admits exponentially decaying solutions away from the interface. 
For semi-infinite material domains, imposing boundedness as $y\to\pm\infty$:
\begin{subequations}
    \label{eq:vel_solution}%
    \begin{equation}
        V^{(1)}_y = A^{(1)} e^{ky} + B^{(1)} e^{q^{(1)} y}, \qquad y < y_{\Gamma}(x,0),
    \end{equation}
    \begin{equation}
        V^{(2)}_y = A^{(2)} e^{-ky} + B^{(2)} e^{-q^{(2)} y}, \qquad y > y_{\Gamma}(x,0),
    \end{equation}
\end{subequations}%
where 
\begin{equation*}
    (q^{(j)})^2 = k^2 + \frac{\rho_{\text{s}}^{(j)} \gamma}{\hat \mu^{(j)}},
\end{equation*}
and coefficients $A^{(j)}$ and $B^{(j)}$ are determined by the interface conditions.
The $e^{\pm k y}$ solutions constitute the irrotational potential mode, whereas the $e^{\pm q^{(j)}y}$ solutions corresponds to the rotational shear branch.
The latter corresponds to the classical treatment of viscous interfacial waves~\cite{Prosperetti1976}. 
The boundedness condition requires $\Real(q^{(j)})>0$, so that the rotational shear-diffusion mode decays away from the interface and remains localized at the interface.
The pressure amplitude and stress components are obtained by substituting \cref{eq:vel_solution} into the linearized momentum equations \cref{eq:normal_momentum} and the constitutive relation \cref{eq:constitutive_law}:
\begin{subequations}
    \begin{equation}
        P^{(j)} = \frac{-\rho_{\text{s}}^{(j)} \gamma D V_y^{(j)} + \hat\mu^{(j)} (D^3V_y^{(j)} - k^2 DV_y^{(j)})}{k^2},
    \label{eq:P_solution}
    \end{equation}
    \begin{equation}
        [\boldsymbol{T}]_{xy}^{(j)} = \hat{\mu}^{(j)} (D V_x^{(j)} + ikV_y^{(j)}), 
    \end{equation}
    \begin{equation}
        [\boldsymbol{T}]_{yy}^{(j)} = 2\hat{\mu}^{(j)} DV_y^{(j)}.
    \end{equation}
\end{subequations}
Then, \cref{eq:interface_conditions} at the interface $\mathbf{x}_{\Gamma}$ becomes:
\begin{subequations}
    \label{eq:interface_normal}
    \begin{equation}
        V_y^{(1)} = V_y^{(2)} = V_y,
    \end{equation}
    \begin{equation}
        V_x^{(1)} = V_x^{(2)} = V_x,
    \end{equation}
    \begin{equation}
        \begin{aligned}
            &k\left[\hat{\mu}^{(1)}\left(q^{(1)}-k\right)-\hat{\mu}^{(2)}\left(q^{(2)}-k\right)\right]V_x+\\
            &\left[\hat{\mu}^{(1)}q^{(1)}\left(k+q^{(1)}\right)+\hat{\mu}^{(2)}q^{(2)}\left(k+q^{(2)}\right)\right]V_y=0,
        \end{aligned}
    \label{eq:normal_stress_condition}
    \end{equation}
    \begin{equation}        
        \left[\hat{\mu}^{(1)}\left(k+q^{(1)}\right)+\hat{\mu}^{(2)}\left(k+q^{(2)}\right)\right]V_x+i\left[\hat{\mu}^{(1)}\left(k-q^{(1)}\right)-\hat{\mu}^{(2)}\left(k-q^{(2)}\right)\right]V_y=0.
        \label{eq:tangential_stress_condition}
    \end{equation}
\end{subequations}

\begin{figure}
    \centering
    \includegraphics[width=\linewidth]{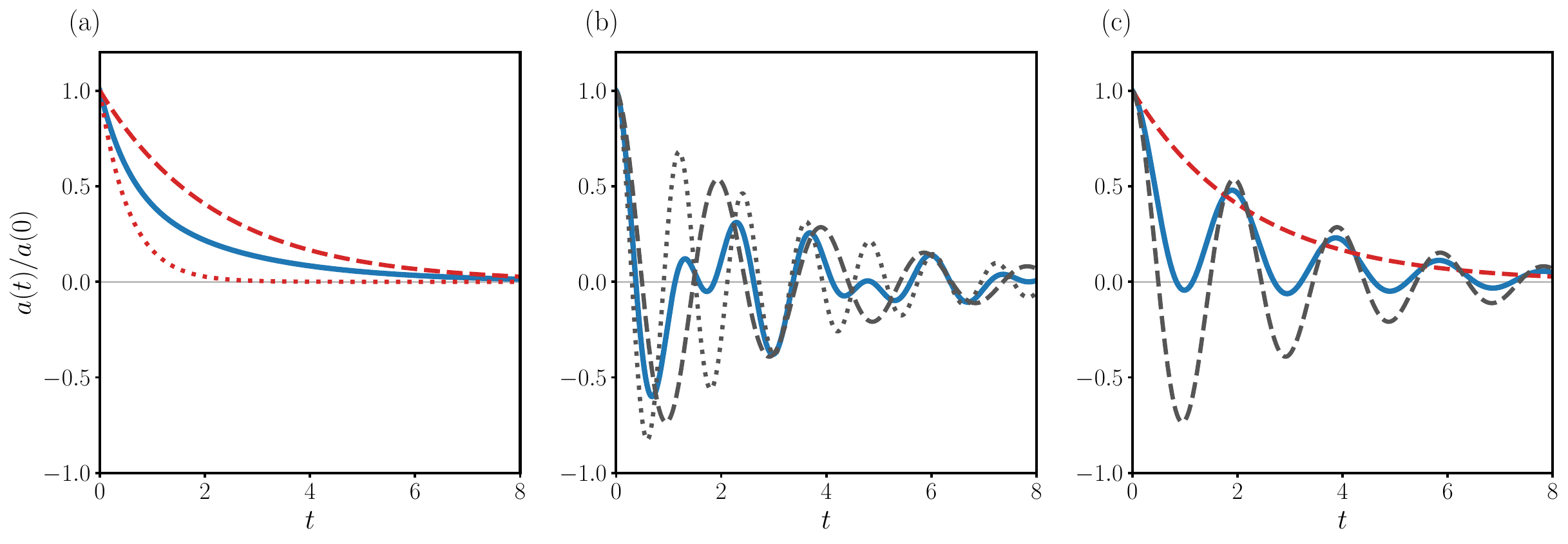}
    \caption{Interface amplitude responses versus time. 
    (a) two real growth rates; (b) two complex-conjugate pairs; (c) mixed real and complex-conjugate growth rates. 
    Red curves: real growth rates; gray curves: oscillatory complex-conjugate pairs; blue solid: equally weight combined interface responses.
    Dashed and dotted curves are high and low rates or frequency, respectively.
    }
    \label{fig:model_response}
\end{figure}

Substituting \cref{eq:vel_solution} into \cref{eq:interface_normal} yields the homogeneous linear system for the coefficients $A^{(j)}$ and $B^{(j)}$: $\mathbf{M}\mathbf{x} = \mathbf{0}$, where
\begin{equation}
    \mathbf{M} =
    \begin{bmatrix}
        1 & 1 & -1 & -1\\
        k & q^{(1)} & k & q^{(2)}\\
        \alpha_1 & \beta_1 & -\alpha_2 & -\beta_2\\
        k\beta_1 & q^{(1)}\alpha_1 & k\beta_2 & q^{(2)}\alpha_2
    \end{bmatrix},
    \qquad
    \mathbf{x} = 
        \begin{bmatrix} 
            A^{(1)}\\ B^{(1)}\\ A^{(2)}\\ B^{(2)}
        \end{bmatrix},
    \label{eq:matrix}
\end{equation}
with $\alpha_j = 2\hat \mu^{(j)} k^2$ and $\beta_j = 2\hat \mu^{(j)} k^2+\rho_{\text{s}}^{(j)} \gamma$.
Nontrivial solutions exist for \cref{eq:matrix} when
\begin{equation}
\begin{aligned}
    \det (\mathbf{M}) = &(q^{(1)}-k)(q^{(2)}-k)\,k^2 \left[\gamma(\rho_{\text{s}}^{(1)} +\rho_{\text{s}}^{(2)} ) \left( \hat \mu^{(1)} +\hat \mu^{(2)}\frac{q^{(2)}}{k} + \hat \mu^{(2)}+\hat \mu^{(1)}\frac{q^{(1)}}{k} \right) \right.\\
    & \left. + 4k^2\left(\hat \mu^{(1)}+\hat \mu^{(2)}\frac{q^{(2)}}{k}\right)\left(\hat \mu^{(2)}+\hat \mu^{(1)}\frac{q^{(1)}}{k}\right)
    \right] = 0.
\end{aligned}    
\label{eq:appendix_det_factorized}
\end{equation}
The factors $(q^{(j)}-k)$ correspond to the zero-growth-rate branch. 
The nontrivial post-shock instability growth rate is in the square brackets and, after algebraic manipulation, the dispersion relation is
\begin{equation}
    \mathcal{D}:=\gamma\left[\frac{1}{\hat{\mu}^{(1)}+\hat{\mu}^{(2)}q^{(2)}/k}+\frac{1}{\hat{\mu}^{(2)}+\hat{\mu}^{(1)}q^{(1)}/k}\right]+\frac{4k^2}{\rho_{\text{s}}^{(1)}+\rho_{\text{s}}^{(2)}}=0,
    \label{eq:dispersion_q}
\end{equation}
or equivalently,
\begin{equation}
    \mathcal{D}:= \gamma\left[\frac{1}{\hat{\mu}^{(1)}+\hat{\mu}^{(2)}\sqrt{1+\dfrac{\rho_{\text{s}}^{(2)}\gamma}{\hat{\mu}^{(2)} k^2}}}+\frac{1}{\hat{\mu}^{(2)}+\hat{\mu}^{(1)}\sqrt{1+\dfrac{\rho_{\text{s}}^{(1)}\gamma}{\hat{\mu}^{(1)} k^2}}}\right] +\frac{4k^2}{\rho_{\text{s}}^{(1)}+\rho_{\text{s}}^{(2)}} = 0.
    \label{eq:dispersion_gamma}
\end{equation}
This equation generally admits $N_p$ number of growth rate $\gamma_p$ indexed by $p$. 
The interface amplitude can then expressed as
\begin{equation}
    a(t)= \sum_p^{N_p} c_p e^{\gamma_p t},
    \label{eq:interface_evolution}
\end{equation}
where the coefficients $c_p$ are determined by the post-shock initial conditions.
Each growth rate can be expressed in complex form: $\gamma_p = \Real(\gamma_p) + i \operatorname{Im}(\gamma_p)$.
Accordingly, \Cref{fig:model_response} shows three of the four distinct interface responses corresponding to these growth rates.
The first type of response involves a solution with only real growth rates ($\operatorname{Im}(\gamma_p)=0$) and the perturbation growth/decay rate is monotonic (see \cref{fig:model_response}(a)).
In the second type, complex-conjugate growth rate pairs ($\operatorname{Im}(\gamma_p)\neq0$) and oscillatory motion with frequency ($|\operatorname{Im}(\gamma_p)|$) of the interface is exhibited (see \cref{fig:model_response}(b)). 
Both real and complex-conjugate pair growth rates are present in the third type.
This interface response consists of a superposition of monotonic and oscillatory components (see \cref{fig:model_response}(c)).
For the fourth there are no admissible growth rates (not shown).

\subsection{Nondimensionalization \label{sec:nondimensional}}

We first examine two limiting cases of the dispersion relation corresponding to purely viscous and purely elastic responses.
For the symmetric case
$\rho_{\mathrm{s}}^{(1)}=\rho_{\mathrm{s}}^{(2)}=\rho_{\mathrm{s}}$,
$\mu^{(1)}=\mu^{(2)}=\mu$,
and
$G^{(1)}=G^{(2)}=G$,
the purely viscous limit ($G^{(j)}\rightarrow0$) and the purely elastic limit ($\mu^{(j)}\rightarrow0$) give
$\hat{\mu}^{(j)}=\mu$
and
$\hat{\mu}^{(j)}=G/\gamma$,
respectively.
The corresponding viscous and elastic dispersion relations are
\begin{equation}
    \mathcal{D}:=\gamma \left[\frac{2}{\mu\left(1+\sqrt{1+\frac{\rho_{\text{s}}\gamma}{\mu k^2}}\right)}\right]+\frac{2k^2}{\rho_{\text{s}}} =0,
\end{equation}
and
\begin{equation}
    \mathcal{D}:=\gamma^2 \left[ \frac{2}{G \left(1+\sqrt{1+\frac{\rho_{\text{s}}\gamma^2}{G k^2}}\right)}\right]+\frac{2k^2}{\rho_{\text{s}}} =0,
\end{equation}
with nontrivial, characteristic solutions
\begin{equation}
    \gamma_{\mathrm{v}}
    =-\frac{\mu k^2}{\rho_{\mathrm{s}}}, \qquad
    \gamma_{\mathrm{e}}
    =\pm ik\sqrt{\frac{G}{\rho_{\mathrm{s}}},
}
\end{equation}
respectively.

We define the viscous and elastic characteristic response rates for the general two-material configuration as
\begin{equation}
    \gamma^{\tv} = \frac{\mu_{\mix} k^2}{\rho_{\mix}}, \qquad
    \gamma^{\text{e}} = k\sqrt{\frac{G_{\mix}}{\rho_{\mix}}},
    \label{eq:characteristic_rates}
\end{equation}
respectively, where $\phi_{\mix}$ denotes the sum of a material property: $\phi_{\mix} = \phi^{(1)} + \phi^{(2)}$.
The elastic growth rate is the magnitude of the elastic oscillation frequency.
The viscous growth rate scales quadratically with the perturbation wavenumber, whereas the elastic response rate scales linearly.
Consequently, long-wavelength perturbations ($k \ll 1$) tend to favor oscillatory responses, while short-wavelength perturbations ($k \gg 1$) tend to exhibit more strongly damped, monotonic behavior.

The dispersion relation is nondimensionalized using the wavenumber $k$ and the viscosity-dominated growth rate for space and time, respectively, and the following nondimensional groups.
We define the competition of the elasticity- and viscosity-dominated growth rates as the viscoelastic ratio:
\begin{equation}
    \Gamma = \frac{\gamma^{\tv}}{\gamma^{\text{e}}},
\end{equation}
and it is regularized as a viscoelastic parameter:
\begin{equation}
    \Ev=\frac{\Gamma - 1}{\Gamma + 1}.
    \label{eq:viscoelastic}
\end{equation}
$\Ev \to 1$ and $\Ev \to -1$ correspond to the purely viscous and elastic limits, respectively.
The material density asymmetry across the interface is quantified with the Atwood number:
\begin{equation}
    \A=\frac{\rho_{\text{s}}^{(1)}-\rho_{\text{s}}^{(2)}}{\rho_{\mix}},
\end{equation}
and the respective Atwood-like numbers for the viscosity and shear modulus contrasts:
\begin{equation}
    \A_\mu=\frac{\mu^{(1)}-\mu^{(2)}}{\mu_{\mix}}, \qquad
    \A_G=\frac{G^{(1)}-G^{(2)}}{G_{\mix}}. 
\end{equation}
The nondimensional dispersion relation (\cref{eq:dispersion_gamma}) is then
\begin{equation}
    \mathcal D := s\Bigg[\frac{1}{\chi_++\chi_- \xi_-}+ \frac{1}{\chi_- +\chi_+ \xi_+}\Bigg] +2 = 0
    \label{eq:dispersion_nondimensional}
\end{equation}
where
\begin{equation*}
    \xi = \frac{q}{k}, \quad 
    s = \frac{\gamma}{\gamma^{\tv}}, \quad
    \xi_\pm=\sqrt{1+\frac{(1\pm \A)s}{\chi_\pm}}, \quad
    \chi_\pm=(1\pm \A_\mu)+\frac{(1-\Ev)^2}{s(1+\Ev)^2}(1\pm \A_G).
\end{equation*}
$s$ is the nondimensional growth rate and $\chi_\pm$ auxiliary variables simplifying notation.
For a material pair, $k$ is the only independent parameter of the system via $\Gamma$.

\subsection{Solution procedure of the nondimensional dispersion relation}
\label{app:sol_procedure}

\Cref{eq:dispersion_nondimensional} contains multiple square-root terms and is nonlinear in $s$. 
It can be written as a 14-order polynomial equation:
\begin{equation}
    P_{14}(s;\Ev,\A,\A_\mu,\A_G)=0.
\label{eq:polynomial_dispersion}
\end{equation}
The transformation is constructed such that every solution of the original dispersion relation is also a root of $P_{14}$. 
The polynomial reformulation does not lead to root loss, but introduces additional candidate roots.
Each candidate root is required to satisfy the domain of the dispersion relation, i.e.,
\begin{equation}
    \begin{aligned}
        & s \neq 0, \qquad \chi_+(s)\neq0, \qquad \chi_-(s)\neq0,\\
        &\chi_+(s)+\chi_-(s)\xi_-(s)\neq0, \qquad \chi_-(s)+\chi_+(s)\xi_+(s) \neq 0.
    \end{aligned}
\end{equation}
For every domain-valid candidate, the residual of \cref{eq:dispersion_nondimensional} is evaluated. 
Only roots satisfying
\begin{equation}
    \left|\mathcal D(s;\Ev,\A,\A_\mu,\A_G)\right|<\varepsilon,
\end{equation}
where $\varepsilon = \SI{E-10}{}$ is a small numerical tolerance, and the far-field decay boundary conditions $\Real(\xi^{(1)})>0$ and $\Real(\xi^{(2)})>0$ are retained as admissible growth rates of \cref{eq:dispersion_nondimensional}. 

\section{Results and discussion\label{sec:results}}

\subsection{Temporal growth rate responses}
\begin{table}[!b]
\centering
\caption{Material parameter cases for the different growth rate stability studies.}
\label{tab:material_configurations}
\small
\renewcommand{\arraystretch}{1.3}
\setlength{\tabcolsep}{5pt}
\begin{tabular}{l c c l}
\hline
Case & $\A$ & $\A_\mu$ & Example\\
\hline
Baseline: Identical density and viscosity & $0$ & $0$ & ---\\
Low Atwood number, identical viscosity & $-0.05$ & $0$ & Hydrogel--Hydrogel \\
Low Atwood number, high viscosity-contrast & $-0.05$ & $\approx -1$ & Water--Hydrogel \\
High Atwood number, high viscosity-contrast &$\approx -1$ &$\approx -1$ & Foam--Hydrogel\\
\hline
\end{tabular}
\end{table}

\Cref{tab:material_configurations} tabulates the four different material parameter cases in which $\A$ and $\A_\mu$ are fixed and $\Ev$ and $\A_G$ are varied between $-1$ to $1$.
The baseline case is a symmetric material with zero $\A$ and $\A_\mu$ with a closed-form solution.
Since a non-zero Atwood number is required for shock-driven interface instability through baroclinic vorticity deposition, we consider two hydrogels with different densities. 
Typical hydrogels have densities ranging from $\SIrange{E3}{1.1E3}{\kilo\gram\per\meter\cubed}$~\cite{Dong2020,Ahart2023}, thus we use $\A=-0.05$. 
We then consider an interface between a low-viscosity liquid and highly viscous viscoelastic material, e.g., a water--hydrogel system with $\A=-0.05$ and $\A_\mu\approx-1$. 
For this case, the viscosity of water is $\mu_{\text{water}} = \SI{E-3}{\pascal\second}$ and ranges from $\SIrange{10}{E2}{\pascal\second}$ for hydrogels depending on composition and water content \cite{Deng2025}.
We then consider materials with both high Atwood number and high viscosity-contrast, e.g., foam–hydrogel, hydrogel–elastomer.
At room temperature ($\SI{20}{\celsius}$) and atmospheric pressure,  $\rho_{\rm air}\approx \SI{1.2}{\kilo\gram\per\meter\cubed}$ and $\mu_{\text{air}}\approx\SI{18}{\micro\pascal\second}$.
Thus, $\A\approx -1$ and $\A_\mu\approx -1$ are used for an air and hydrogel example.

\subsubsection{Baseline case \label{sec:results_symmetric}}

Although there is no material contrast, $\Ev$ is still a function of the viscosity and elasticity of the two materials and \cref{eq:dispersion_nondimensional} reduces to
\begin{equation}
    \mathcal{D}(s;\Ev) = s^2 + s + \left( \frac{1 - \Ev}{1 + \Ev}\right)^2 = 0,
    \label{eq:dispersion_symmetric}
\end{equation}
where
\begin{equation}
    s_\pm = -\frac{1}{2} \pm \frac{1}{2}\sqrt{1 - 4\left(\dfrac{1 - \Ev}{1 + \Ev}\right)^2}.
\end{equation}
As $\Ev$ increases, the complex-conjugate growth rate pair coalesces into two real growth rates when the discriminant vanishes.
This defines a critical viscoelastic number, denoted by $\Ev_{\tc}$. 
For this material configuration, $\Ev_{\tc}=1/3$, which marks the critical damping threshold separating oscillatory and monotonic interface dynamics.

\begin{figure}[t]
    \centering
    \includegraphics[width=\linewidth]{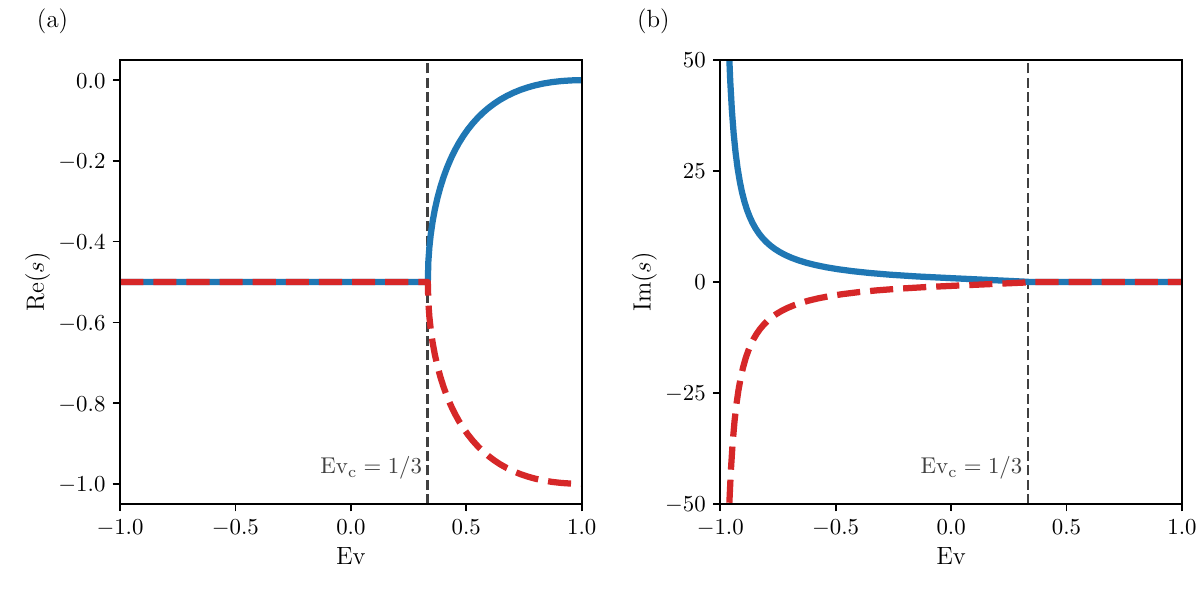}
    \caption{Real (a) and imaginary (b) components of the growth rates versus $\Ev$ for the baseline case.
    Blue solid and red dashed: growth rate branches $s_\pm$, respectively; vertical gray dashed line: $\Ev_{\tc}=1/3$.}
    \label{fig:symmetric_root}
\end{figure}

\Cref{fig:symmetric_root} shows the growth rates versus of $\Ev$. 
Two distinct real growth rates form for $\Ev>\Ev_{\tc}$ and a complex-conjugate pair for $\Ev<\Ev_{\tc}$.
At $\Ev_{\tc}$, the imaginary components of this pair vanishes. 
This transition has an equivalent spring--mass--damper system interpretation.
Considering the dimensional form of \cref{eq:dispersion_symmetric} with \cref{eq:interface_evolution} gives
\begin{equation}
    \ddot a+\gamma^{\tv}\dot a+(\gamma^{\text{e}})^2a=0.
\end{equation}
This is a damped second-order dynamical system governing the interface amplitude.
The corresponding critical-damping condition is $\gamma^{\tv}=2\gamma^{\text{e}}$ and $\Ev_{\tc}=1/3$.
The complex-conjugate growth rate pair shown corresponds to an underdamped interface response, whereas the two real growth rates correspond to an overdamped response. 

\begin{figure}[t]
    \centering
    \includegraphics[width=0.8\linewidth]{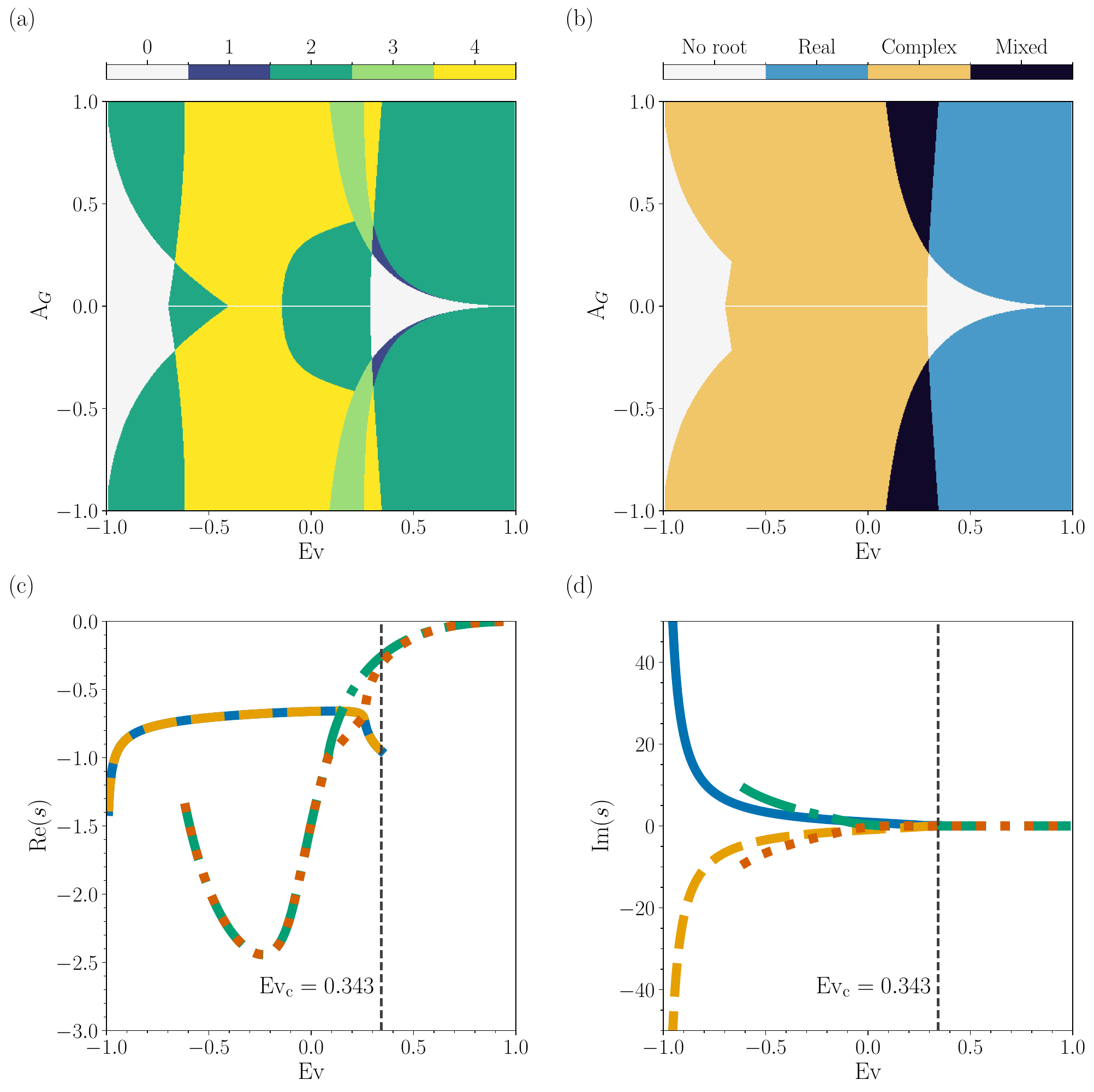}
    \caption{Color contour of the growth rate (a) count and (b) regime versus $\A_G$ and $\Ev$ for the baseline case.
    Real (c) and imaginary (d) components of the growth rates for $\A_G=-1$. 
    Blue solid, orange dashed, green dash-dotted, and red dotted curves: different growth rates; vertical gray dashed line: $\Ev_{\tc}$.}
    \label{fig:2Dmap_0Rho_0Mu}
\end{figure}

\Cref{fig:2Dmap_0Rho_0Mu} shows the growth rate count and regime contours with $-1\le \A_G\le1$ (a, b) and the real and the imaginary growth rates for $\A_G=-1$ (c, d) for the baseline case.
The two materials are identical except for their elastic moduli.
Thus, reversing the sign of $\A_G$ is equivalent to interchanging the two materials and does not change the governing equations, the interface conditions, or the dispersion relation. 
The resulting growth rate regimes are symmetric about $\A_G=0$.
The zero to four growth rates form four distinct regimes: no-growth-rate, pure-real, pure-complex, and mixed. 
The mixed regime forms a narrow transition band that separates the pure-complex and pure-real regimes (see \Cref{fig:2Dmap_0Rho_0Mu}(a,b)).
Increasing $\Ev$ shifts the growth rate regime from complex to real and $\A_G$ determines the pathway of this transition. 
For $|\A_G|<0.26$, the transition takes place between the pure-complex and pure-real regimes through an intervening no-growth-rate region.
As $|\A_G|$ increases, the no-growth-rate regime shrinks and becomes mixed.
When $\A_{G} = -1$, the critical viscoelastic parameter shifts from $\Ev_{\tc}=1/3$ to the more viscous $\Ev_{\tc}=0.343$ (see \Cref{fig:2Dmap_0Rho_0Mu}(c,d)). 
As $\Ev \to 1$, viscous damping also vanishes as $\mathrm{Re}(s) \to 0$.

\subsubsection{Low Atwood number, identical viscosity case \label{sec:results_weak_weak}}

\begin{figure}[t]
    \centering
    \includegraphics[width=\linewidth]{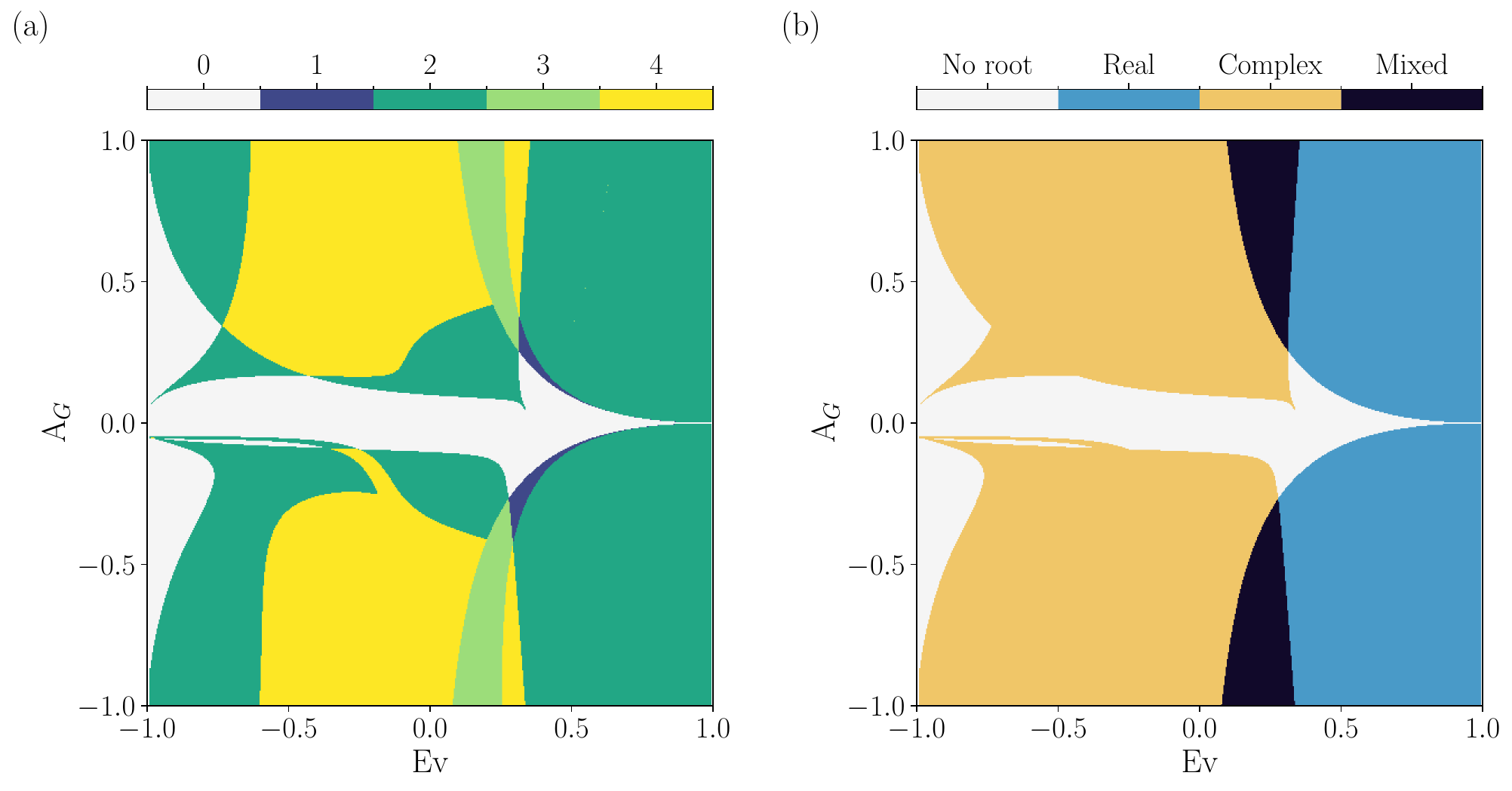}
    \caption{Color contour of the growth rate (a) count and (b) regime versus $\A_G$ and $\Ev$ for the low Atwood number, identical viscosity case.
    }
    \label{fig:2Dmap_LowRho_LowMu}
\end{figure}
\begin{figure}[t]
    \centering
    \includegraphics[width=0.75\linewidth]{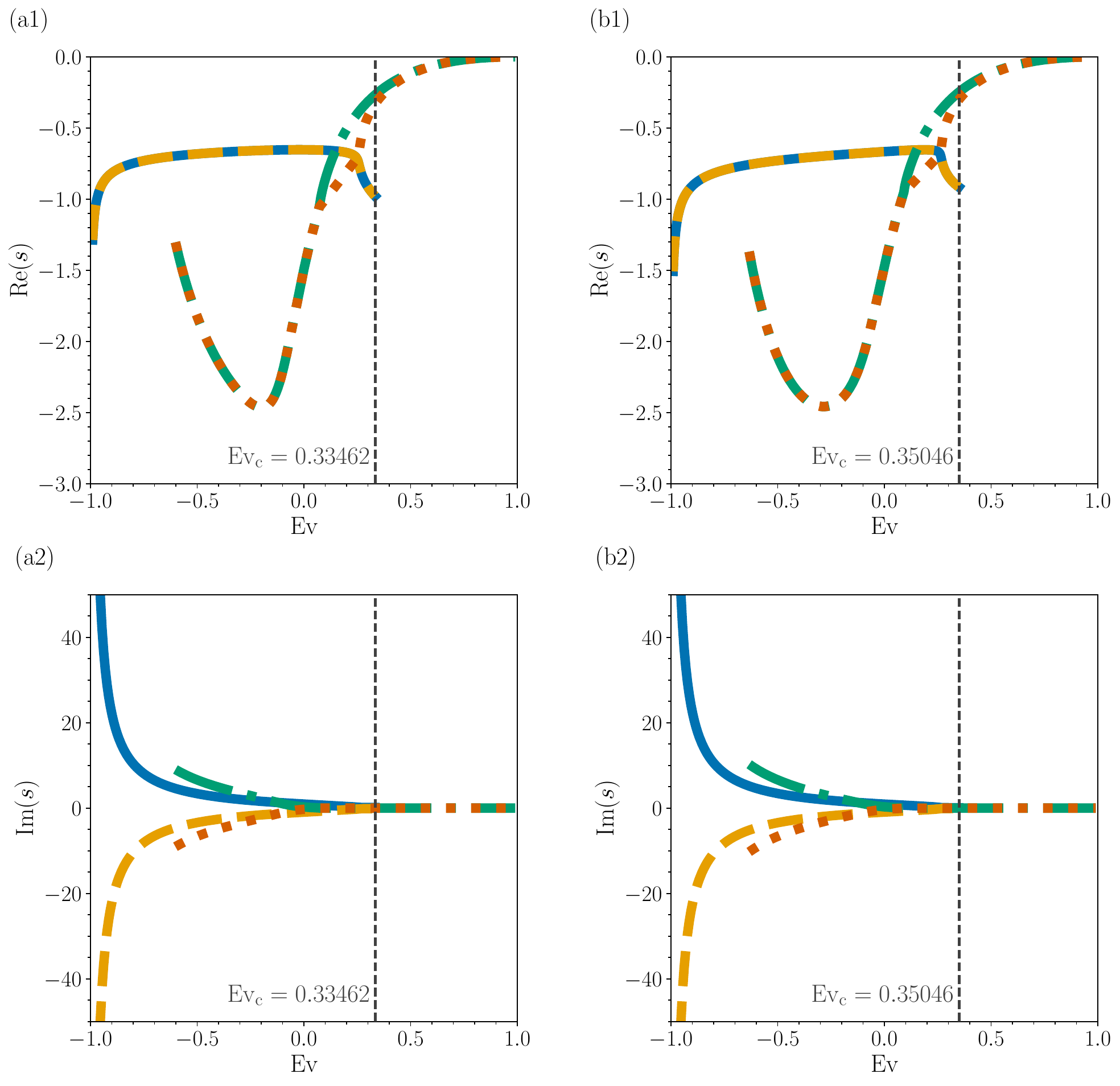}
    \caption{Real (row 1) and imaginary (row 2) components of the growth rates for the low Atwood number and identical viscosity case with (a) $\A_G=-1$ and (b) $\A_G=1$. 
    Blue solid, orange dashed, green dash-dotted, and red dotted curves: different growth rates; vertical gray dashed line: $\Ev_{\tc}$.
    }
    \label{fig:eigenvalues_LowRho_LowMu}
\end{figure}

A low, finite Atwood number breaks the symmetry of the growth rate regime depending on the value of $\A_G$ by shifting the transition $\Ev$ values between regimes (see \Cref{fig:2Dmap_LowRho_LowMu}). 
Compared with the symmetric material case, the topological change is the expansion of the no-growth-rate region around $\A_G\approx0$ for $\Ev<0$. 
The shifting of $\Ev_{\tc}$ depends on the sign of the product of $\A$ and $\A_G$ (see \Cref{fig:eigenvalues_LowRho_LowMu}).
For $\A_G=-1$, the critical viscoelastic parameter is $\Ev_{\tc}=0.335$, which is less than the symmetric case of $\Ev_{\tc}=0.343$. 
Since the denser material is also the stiffer material (i.e., $\A$ and $\A_G$ have the same sign), oscillatory interface responses can be suppressed under weaker viscosity dominance and is reflected by a smaller $\Ev_{\tc}$.
On the other hand, the critical viscoelastic parameter increases to $\Ev_{\tc}=0.350$ for $\A_G=1$.
In this scenario, the denser material is more elastically compliant, and leads to a larger $\Ev_{\tc}$ relative to the symmetric case.

\subsubsection{Low Atwood number, high viscosity-contrast case \label{sec:results_weak_strong}}
\begin{figure}[t]
    \centering
    \includegraphics[width=\linewidth]{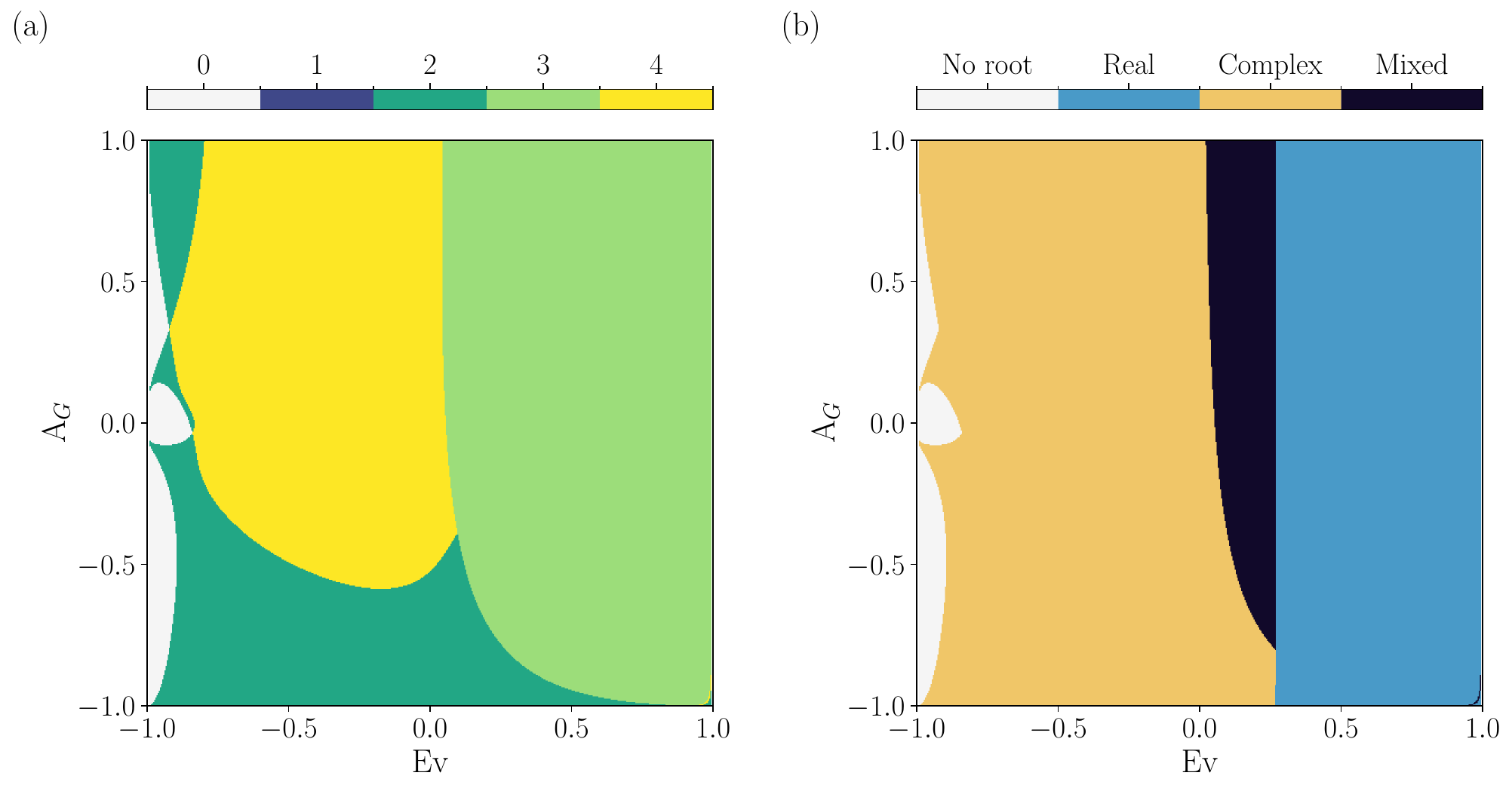}
    \caption{Color contour of the growth rate (a) count and (b) regime versus $\A_G$ and $\Ev$ for low Atwood number, high viscosity-contrast case.}
    \label{fig:2Dmap_LowRho_HighMu}
\end{figure}

\Cref{fig:2Dmap_LowRho_HighMu} shows the contours of the growth rate count and regime versus $\A_G$ and $\Ev$ for the low Atwood number and high viscosity-contrast case.
The mixed growth rate regime progressively narrows as $\A_G$ decreases and disappears at $\A_G\lesssim-0.8$, resulting in a direct transition from pure-complex to pure-real regime.
The no-growth-rate region is confined to a much smaller parameter range near $\Ev\approx-1$.
In particular, along $\A_G=0$, nontrivial growth rates appear for $\Ev\gtrsim-0.85$.

\begin{figure}[t]
    \centering
    \includegraphics[width=\linewidth]{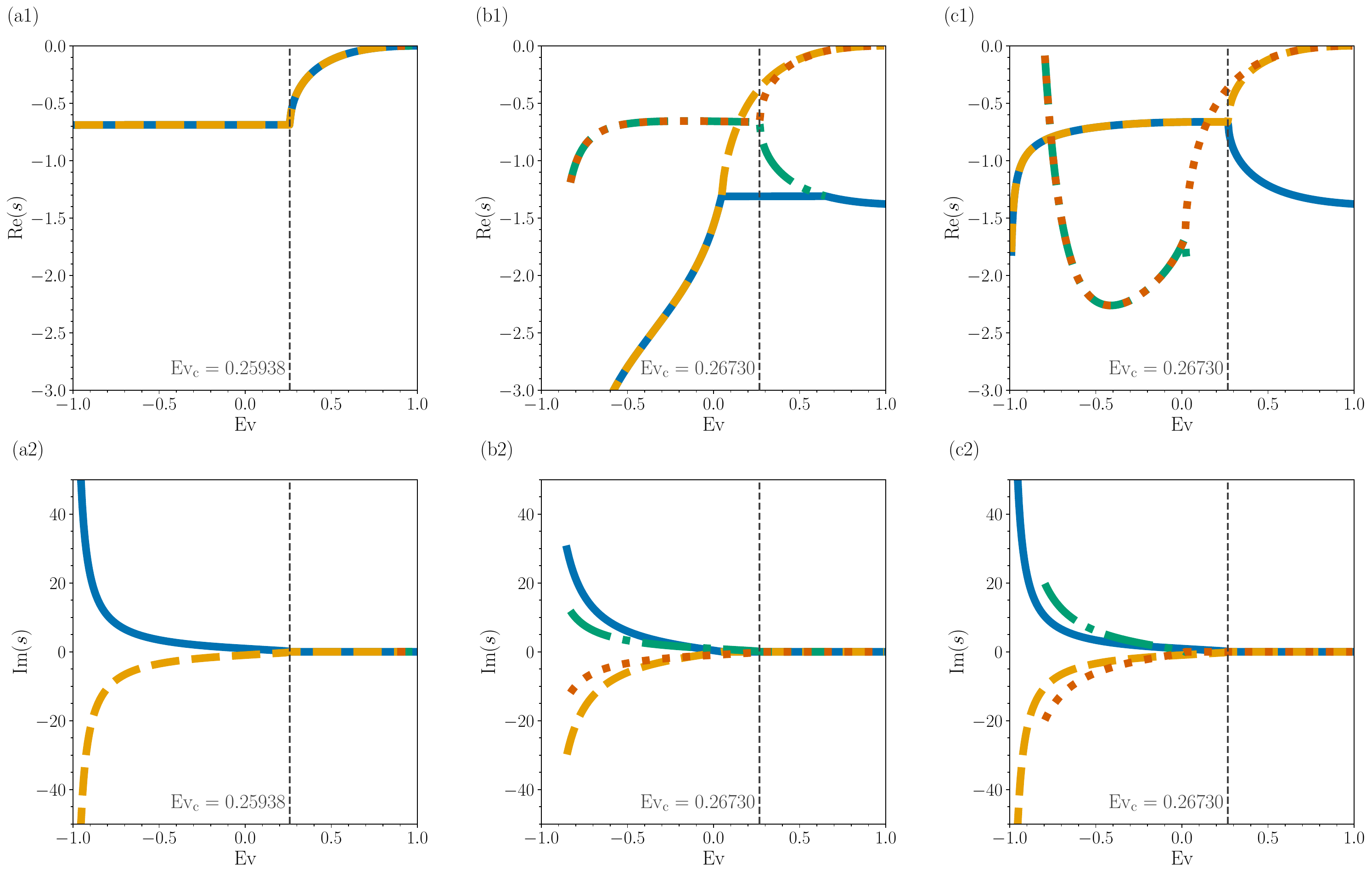}
    \caption{Real (row 1) and imaginary (row 2) temporal growth rates for the low Atwood number, high viscosity-contrast case. 
    (a) $\A_G=-1$; (b) $\A_G=0$; (c) $\A_G=1$.
    Blue solid, orange dashed, green dash-dotted, and red dotted curves: different growth rates; vertical gray dashed line: $\Ev_{\tc}$.}
    \label{fig:eigenvalues_LowRho_HighMu}
\end{figure}

Relative to the identical viscosity ($\A_\mu = 0$) case (see \cref{fig:eigenvalues_LowRho_LowMu}), the pure-real growth rate regime extends into smaller $\Ev$ values reflected by the smaller $\Ev_c$ values (see \Cref{fig:eigenvalues_LowRho_HighMu}).
As expected, depending on value of $\A_G$, the two Kelvin–Voigt material model configuration becomes asymmetric in its response. 
For $\A_G=-1$, the upper material is elastically stiffer and viscous, the transition is followed by two close real growth rates with increasing $\Ev$ that vanish as $\Ev \to 1$. 
The aligned signs of $\A_{G}$ and $\A_{\mu}$ suppress additional growth rate branches in the system.
As $\A_G\to1$, the elasticity and viscosity contrasts are oppositely oriented. 
As $\Ev$ increases, two of the three growth rates approach $\mathrm{Re}(s)=0$ and the third has a finite-rate real value. 
The identical shear modulus case ($\A_G=0$) exhibits behavior between these two limiting cases.
Additionally, $\Ev_c$ weakly depends on $\A_G$ in this case with a variation less than $0.01$.

\subsubsection{High Atwood number, viscosity-contrast case \label{sec:results_strong_strong}}

\begin{figure}[t]
    \centering
    \includegraphics[width=\linewidth]{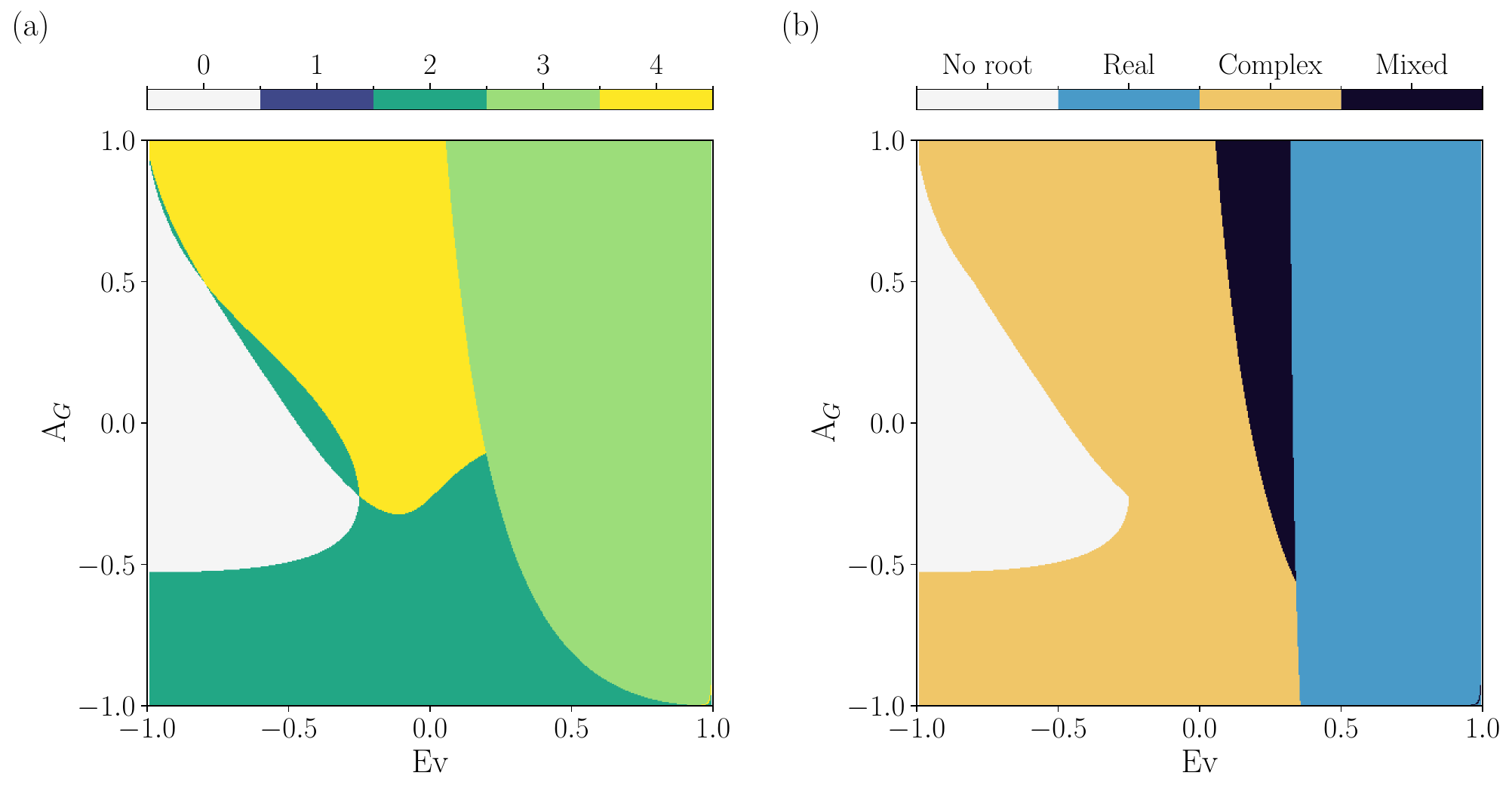}
    \caption{Color contour of the (a) count and (b) regime of the growth rates versus $\A_G$ and $\Ev$ space for high Atwood number, high viscosity-contrast case.
    }
    \label{fig:2Dmap_highRho_highMu}
\end{figure}

Relative to the low Atwood number, high $\A_{\mu}$ case (\cref{fig:2Dmap_LowRho_HighMu}), \Cref{fig:2Dmap_highRho_highMu} shows the transition between pure-real growth rate and complex and mixed regimes shifts at larger $\Ev$ values. 
Additionally, the mixed growth rate regime is compressed. 
For $\A_G<0$, the mixed regime is restricted to $\A_G\gtrsim-0.56$, beyond which the transition proceeds directly from complex-dominated to purely real growth rate.
Notably, the no-growth-rate region expands, particularly for $\A_G<0$. 
Increasing $\Ev$ continues to drive the growth rate structure from complex-dominated to real-dominated regimes.

\begin{figure}[t]
    \centering
    \includegraphics[width=\linewidth]{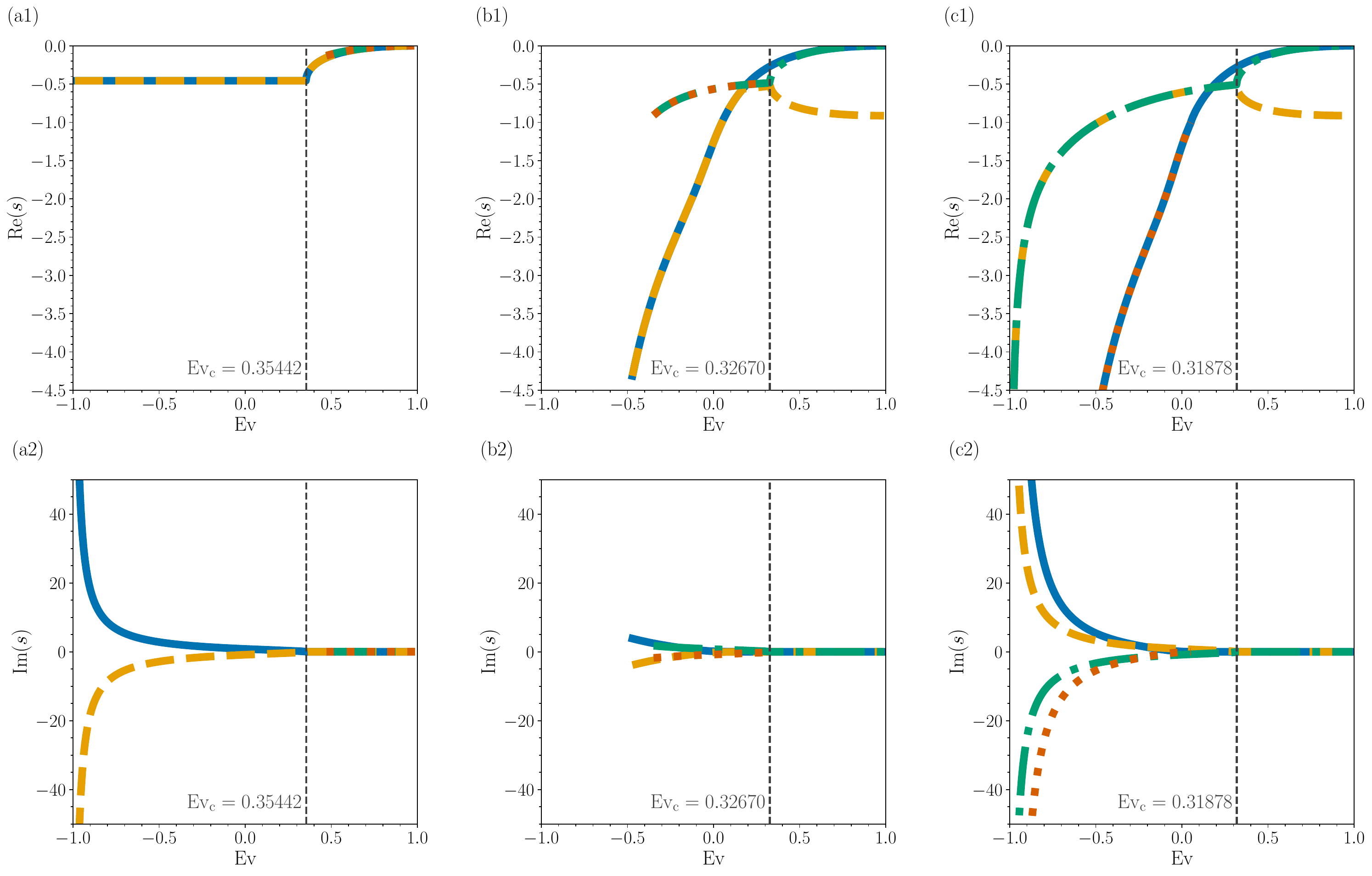}
    \caption{Real (row 1) and imaginary (row 2) growth rates for the high Atwood number, high viscosity-contrast case. 
    (a) $\A_G=-1$; (b) $\A_G=0$; (c) $\A_G=1$.
    Blue solid, orange dashed, green dash-dotted, and red dotted curves: different growth rates; vertical gray dashed line: $\Ev_{\tc}$.
    }
    \label{fig:eigenvalues_HighRho_HighMu}
\end{figure}

This case also has the similar overall transition behavior for growth rate regimes to that of the low Atwood number, high $\A_{\mu}$ case.
However, $\Ev_{\tc}$ values are larger than the baseline case and depend on $\A_G$, the total $\Ev_{\tc}$ variation is about $0.036$ over the $\A_G$ domain (see \Cref{fig:eigenvalues_HighRho_HighMu}).
For $\Ev>\Ev_{\tc}$, the real growth rate branches follow similar behaviors as those in low Atwood number, high-viscosity case.
For $\A \approx - 1$, inertia and viscous dissipation are concentrated in the upper material, resulting in an one-sided dynamical system.
As a result, the overall dissipation of the two-material system is reduced and less able to suppress oscillatory motion produced by the upper material elasticity.
Thus, $\Ev_{\tc}$ must be closer to unity is to eliminate the oscillatory modes with $\A_{G}$ controlling the magnitude of this shift.
For $\A_G < 0$, elasticity is higher in the upper, denser, more viscous material.
The elastic contribution of the stress (see~\cref{eq:cauchy_viscous}) of this upper material carries the elastic energy and inertia which allows oscillatory modes to persist and increases $\Ev_{\tc}$.
In contrast when $\A_G \ge 0$, the lower material is elastically stiffer and weakens the elastic-inertial coupling of the upper material to yield lower $\Ev_{\tc}$ values than the baseline case.

\subsection{Material dependence of the viscoelastic transition \label{sec:critical_Ev}}

\begin{figure}[t]
    \centering
    \includegraphics[width=0.7\linewidth]{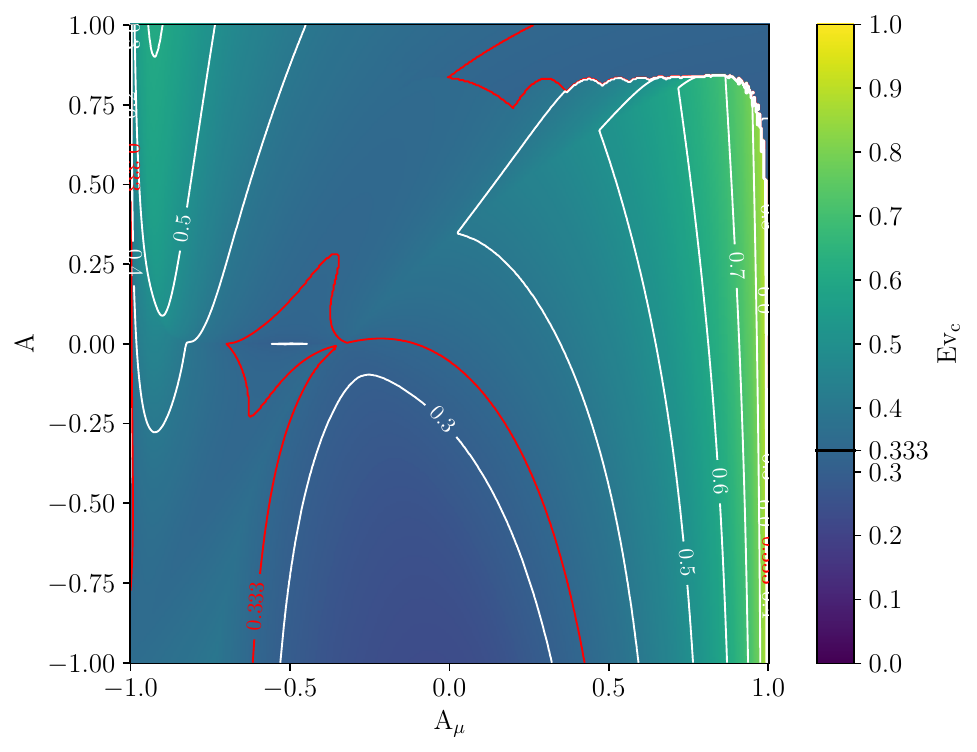}
    \caption{Color contour of $\Ev_{\tc}$ versus $\A$ and $\A_\mu$ with $\A_G=-1$.
    Red contour line: $\Ev_{\tc}=1/3$.}
    \label{fig:transition_ek}
\end{figure}

The case studies show that $\Ev_{\tc}$ depends on $\A$ and $\A_\mu$ and minimally on $\A_G$. 
We therefore consider $\Ev_{\tc}$ versus $\A$ and $\A_\mu$ with $\A_G=-1$ (see \Cref{fig:transition_ek}).
The symmetric reference value of $\Ev_{\tc}=1/3$ highlighted as a red isocontour line.
$\Ev_{\tc}$ remains positive for $\A$ and $\A_\mu$, indicating that the absence of oscillatory growth rates occur only after the viscous characteristic rate exceeds the elastic counterpart.
Moreover, the $\Ev_{\tc}$ remains close to the symmetric reference value of $1/3$ over most of the parameter space.
Pronounced deviations emerge near $|\A_\mu|\approx1$ where $\Ev_{\tc}$ increases rapidly. 

These results provide a criterion for predicting whether a shock-perturbed viscoelastic interface will exhibit oscillatory or monotonic relaxation from its material properties and perturbation wavelength. 
They also provide a theoretical basis for interpreting shock-loading experiments and for guiding the selection and design of viscoelastic materials with targeted early-time interface responses. 
Building on these results, future work will extend the present framework to compressible and nonlinear systems. 
We will also determine the modal amplitudes from the post-shock initial conditions to theoretically predict the interface evolution.
The predictions will be assessed with direct numerical simulations.

\section{Conclusions\label{sec:summary}}

We use the method of normal modes to evaluate the early-time post-shock evolution of a perturbed interface separating two Kelvin–Voigt viscoelastic materials. 
The post-shock problem was formulated as an incompressible interfacial growth rate problem with inertia, viscous dissipation, and elastic resistance. 
The resulting dispersion relation provides the growth rate governing the interface response and reduces the governing parameter space to the Atwood number and Atwood number-like viscosity and shear modulus contrasts, i.e., $\A$, $\A_\mu$, and $\A_G$, respectively, and the viscoelastic parameter, $\Ev$.
The latter parameter primarily governs the growth rate regimes and is a function primarily of the Atwood number and $\A_{\mu}$.
Moreover, its critical value marks the transition from complex-dominated to pure-real growth rate regimes.
As the viscous contribution increases (i.e., $\Ev \to 1$), the balance shifts from elastic-inertial to viscous dynamics, suppressing oscillatory responses and leading to monotonic dampening.
The Atwood number breaks the $\A_{G}$ domain symmetry of growth rate regimes.
Meanwhile, a non-zero $\A_\mu$ redistributes viscous dissipation between the two materials and determines how effectively oscillatory motion is damped.
$\A_G$ governs the balance of restoring stresses across the interface and determines whether real and complex growth rate coexist near the transition.
These results provide a basis for future extensions to nonlinear and compressible viscoelastic systems.

\section{Acknowledgments}
X.Z. acknowledges support from the Hibbitt Postdoctoral Fellowship in the School of Engineering at Brown University.
M.R., Jr. acknowledges support from the U.S. National Science Foundation under Grant No. 2232427.
Anthropic's Claude and OpenAI Codex were used to assist in the Github repository documentation, all outputs were verified multiple times by the authors.
Scientific content, reasoning, and conclusions are from the authors.
Funding agencies were not involved in study design; in the collection, analysis and interpretation of data; in the writing of the report; or in the decision to submit the article for publication.
The views and conclusions contained herein are those of the authors and should not be interpreted as necessarily representing the official policies or endorsements, either expressed or implied, of the funding agencies.

\bibliography{references}

\end{document}